\documentclass[twocolappendix]{emulateapj}

\usepackage{soul,color}

\shorttitle{Rayleigh Taylor Instability}
\shortauthors{Duffell \& MacFadyen}

\begin{document}

\title{Rayleigh-Taylor Instability in a Relativistic Fireball on a Moving Computational Grid}

\author{Paul C. Duffell and Andrew I. MacFadyen}
\affil{Center for Cosmology and Particle Physics, New York University}
\email{pcd233@nyu.edu, macfadyen@nyu.edu}

\begin{abstract}

We numerically calculate the growth and saturation of the Rayleigh-Taylor instability caused by the deceleration of relativistic outflows with Lorentz factor $\Gamma = 10$, $30$, and $100$.  The instability generates turbulence whose scale exhibits strong dependence on Lorentz factor, as only modes with angular size smaller than $1/\Gamma$ can grow.  We develop a simple diagnostic to measure the kinetic energy in turbulent fluctuations, and calculate a ratio of turbulent kinetic energy to thermal energy of ${\epsilon_{RT} = .03}$ in the region affected by the instability.  Although our numerical calculation does not include magnetic fields, we argue that small scale turbulent dynamo amplifies magnetic fields to nearly this same fraction, giving a ratio of magnetic to thermal energy of $\epsilon_B \sim 10^{-2}$, to within a factor of two.  The instability completely disrupts the contact discontinuity between the ejecta and the swept up circumburst medium.  The reverse shock is stable, but is impacted by the Rayleigh-Taylor instability, which strengthens the reverse shock and pushes it away from the forward shock.  The forward shock front is unaffected by the instability, but Rayleigh-Taylor fingers can penetrate of order 10\% of the way into the energetic region behind the shock during the two-shock phase of the explosion.  We calculate afterglow emission from the explosion and find the reverse shock emission peaks at a later time due to its reduced Lorentz factor and modified density and pressure at the shock front.  These calculations are performed using a novel numerical technique that includes a moving computational grid.  The moving grid is essential as it maintains contact discontinuities to high precision and can easily evolve flows with extremely large Lorentz factors.

\end{abstract}


\section{Introduction}
\label{sec:intro}

In the fireball model \citep{pacz,good}, gamma ray bursts are thought to be generated by a hot explosion which expands and compresses itself into a thin ultrarelativistic shell.  Internal collisions of such shells are thought to generate the prompt burst of gamma rays, after which afterglow emission is produced during the further expansion of the shell.  Eventually this shell transfers its energy into a relativistic blastwave propagating into the circumburst medium.  During the transfer process there is an interesting phase of the evolution, during which the shell is unstable to the Rayleigh-Taylor (RT) instability.

The instability requires a density gradient, for example at the interface between two fluids of different density.  In this case, the two fluids are the material contained in the original explosion (the ejecta) and the ambient gas swept up by the explosion (the circumburst medium).  Once the explosion has begun to sweep up a non-negligible amount of mass, two shocks form.  One of the shocks drives its way forward into the ambient gas, and the other back into the ejecta.  Between the two shocks resides the contact discontinuity, marking the separation between ambient gas and ejecta.  It is this contact discontinuity which is unstable.

The instability should have observable signatures.  It has been suggested that relativistic turbulence is at least partially responsible for the prompt emission of gamma rays \citep{l06,kn09}.  Additionally, the instability has a strong impact on the reverse shock, which might be observable in the afterglow radiation from the burst.  The dynamics of RT should be taken into account in order to accurately predict the magnitude and frequency of the so-called ``optical flash" associated with the reverse shock.  Additionally, the instability produces a shear flow which cascades via the Kelvin-Helmholtz instability into turbulence, which can amplify magnetic fields, facilitating synchrotron emission \citep{levA}.  Finally, it has been suggested that RT fingers might propagate all the way up to the forward shock \citep{levA}, possibly altering the fundamental dynamics of the shock, which are generally thought to be governed by the Blandford-McKee solution \citep{bm} in the ultrarelativistic limit.

These considerations motivate the calculation of the growth of the RT instability in the physical context of a relativistic fireball.  The nonrelativistic case was first treated by \cite{c92}, who analytically calculated the linear growth rate, and also explored numerically the linear growth and nonlinear saturation.  \cite{jn96} later performed a two-dimensional magnetohydrodynamics calculation which demonstrated how magnetic fields tend to align themselves along RT fingers.

The relativistic case is relevant for gamma ray bursts, which have Lorentz factors ${\Gamma \gtrsim 100}$, and relativistic effects qualitatively change the physics of the instability.  First, the growth rate is modified by time dilation \citep{w02}.  Secondly, causality arguments dictate which modes grow and saturate during different phases of evolution, analogous to the growth of structure in cosmology.  Additionally, any turbulence being generated is relativistic turbulence, which may have distinct properties from nonrelativistic turbulence \citep{zrakey,radice}.  Finally, while the Sedov-Taylor solution is a fast attractor, the relativistic Blandford-McKee solution is a slow attractor \citep{g00}.  Any deviation from Blandford-McKee due to RT fingers might persist until the shock becomes nonrelativistic.

To extend the nonrelativistic results into the relativistic regime, \cite{levB} performed a stability analysis on the two-shock solution \citep{ns06} and found linear growth rates which could potentially be large enough to impact the forward shock.  \cite{t06} has studied the stabilizing effect of strong magnetic fields in the context of a decelerating photon-rich shell.

In this work, we evolve a relativistic fireball numerically in two dimensions (2D), to calculate its dynamics, and to explore the consequences of this instability to observations of gamma ray bursts and their afterglows.  We study in detail how the qualitative features of turbulence are changed by relativistic effects, and we describe generally how the dynamics of the two-shock system are affected.  To improve code performance, we use a moving numerical mesh, which follows the radial motion of the bulk flow and reduces the advection errors associated with this motion.  Doing this results in a vastly improved calculation of the shock dynamics, and an increased sensitivity to the subtle effects that must be captured.  The code is also able to perform calculations with large Lorentz factors ($\Gamma \gtrsim 100$), which allows us to probe astrophysically relevant regions of parameter space.

\section{Method}
\label{sec:num}
We solve the special relativistic hydrodynamics equations:
\begin{equation}
\partial_{\mu} ( \rho u^{\mu} ) = 0
\end{equation}
\begin{equation}
\partial_{\mu} T^{\mu \nu} = 0
\end{equation}
\begin{equation}
T^{\mu \nu} = ( (\rho + 4P) u^{\mu} u^{\nu} + P g^{\mu \nu} ) 
\label{eqn:stress}
\end{equation}
where $\rho$ is the proper mass density, $P$ is the pressure, and $u$ is the four velocity.  The coefficient of 4 in Equation \ref{eqn:stress} indicates that we are using an adiabatic equation of state with the relativistic adiabatic index of $4/3$.  We use units for which c = 1.  We also evolve a passive scalar field, $X$, according to
\begin{equation}
\partial_{\mu} ( \rho X u^{\mu} ) = 0.
\end{equation}
This allows us to track the growth of the instability, demarking which fluid elements are in the ejecta and which are in the circumburst medium.  We set $X=0$ for the ejecta behind the contact discontinuity, and $X=1$ for the ambient gas in front.

\begin{figure}
\epsscale{1.0}
\plotone{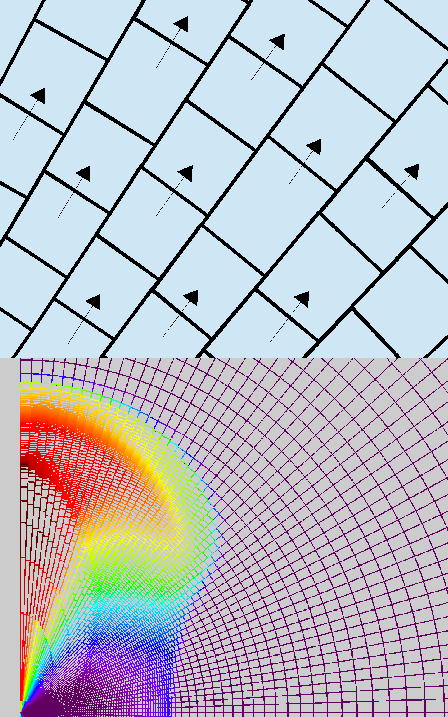}
\caption{ The numerical grid shears with the radial velocity of the gas.  Each radial track moves independently, each behaving essentially as a 1D lagrangian code.  The upper panel shows the motion of grid cells, and the lower panel shows the shearing grid being employed to follow a jet-like explosion.
\label{fig:jetcode} }
\end{figure}

\subsection{Numerics}

The numerical method we employ is a variant of the TESS method \citep{tess} with specific application to problems involving rapid radial outflow.

TESS is a finite-volume, moving-mesh hydrodynamics method which in its original form constructs its numerical mesh from a Voronoi tessellation of the computational domain.  The computational zones move with the local fluid velocity, and change their shape and size as they move and shear past one another.  In this sense, the method is effectively Lagrangian.

This idea can be improved if the moving mesh is not generated from a Voronoi tessellation, but instead is specifically tailored to a given flow.  Most of the numerical benefits from TESS do not come from the Voronoi tessellation, but from the simple fact that cells are allowed to move.  This enables resolution of contact discontinuities to very high precision, and potentially allows for much longer timesteps.  By moving the mesh, but choosing a particular shape for grid cells (other than Voronoi), the code can be adapted to the problem at hand.

The TESS method defines its numerical mesh abstractly, so that all that is required to take into account a given mesh topology are the positions and volumes of computational zones, and the positions and areas of ``faces", which are simply defined to be any boundary between two zones.  This means more specific mesh topologies can be generated without changing anything fundamental about the method.

We have previously made this adaptation for Keplerian flow in a gaseous disk \citep{disco}.  In that case, instead of Voronoi cells, the cells were chosen to be wedge-like annular segments like those of a standard polar grid, and they rotated with the local orbital velocity.  We now specifically refer to the shearing-disk version as the DISCO code.

In this work, we take a similar approach, but instead adapt the grid to a radial outflow, such as a jet.  The wedge-like annular shape is still employed, but instead of rotating, the grid expands and shears radially (Figure \ref{fig:jetcode}).  The numerics are nearly identical to DISCO; we have merely changed the direction of grid motion.  By moving the mesh radially instead of azimuthally, the code effectively behaves as a series of one-dimensional Lagrangian codes, coupled laterally by transverse fluxes.  The moving grid helps to capture the growth of the instability, and it also allows accurate calculations for large Lorentz factors $\Gamma \gtrsim 100$.

We integrate the equations in two dimensions assuming axisymmetry.  The computational domain consists of an angular wedge ${0 < \theta < \pi/16}$.  We use an angular resolution $N_{\theta} = 1600$ and a resolution in the radial dimension of ${N_R \sim 12800}$, which were chosen for a resolution of ${\Delta \theta \sim \Delta r / r \sim 10^{-4}}$.  The number of radial zones varies throughout the course of the calculation, because zones are added and removed from the computational domain as necessary during the course of the calculation.  The aspect ratio of a typical computational zone is of order unity, but if a cell becomes too long or short, the grid is dynamically refined or de-refined by splitting or merging zones.  Aspect ratios in general range between $5$ and $1/5$.  The inner and outer radial boundaries of the domain can also move during the time integration, so that only a fraction of the dynamical range must be covered at any one time.

\subsection{Initial Conditions} 
\label{sec:ics}

\begin{figure*}
\epsscale{1.0}
\plotone{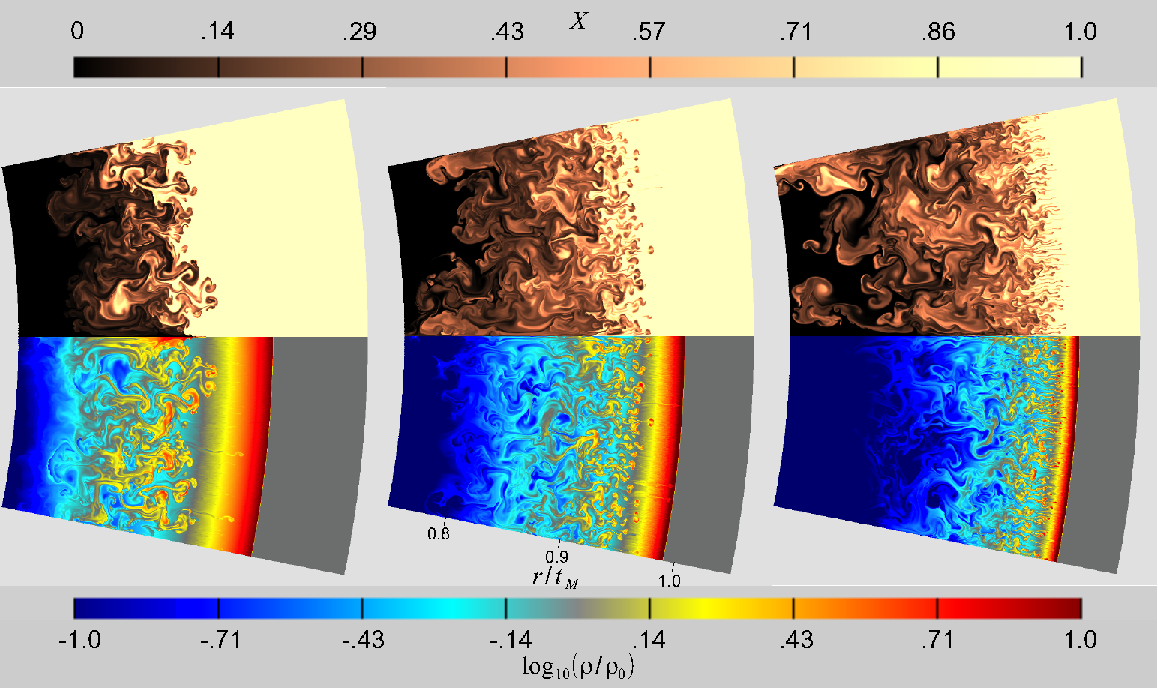}
\caption{ The structure of the turbulence strongly depends on relativistic effects.  When the energy per mass $\Gamma$ is varied (the characteristic Lorentz factor of the initial shell), this affects the scale and prominence of the turbulence generated.  This is a snapshot at $t = t_M$ for $\Gamma = 10, 30,$ and $100$ (left to right).  On the lower part of the figure, we plot logarithm of density, while on the upper part we plot a passive scalar which tracks the growth of the instability.  The dominant mode has angular scale $\Delta \theta \sim 1/\Gamma$.
\label{fig:times} }
\end{figure*}

\begin{figure}
\epsscale{1.0}
\plotone{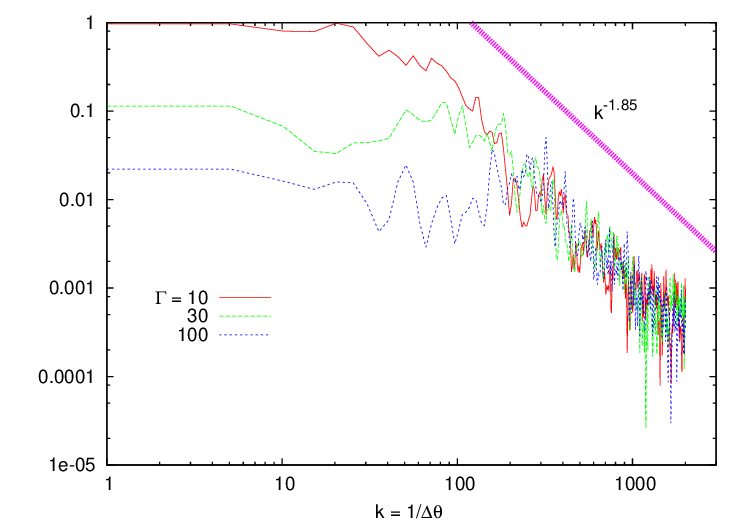}
\caption{ The power spectrum of RT for the three examples shown in Figure \ref{fig:times}.  Curves are calculated by measuring the Fourier transform of $\delta r / r$ in front of the RT instability.  There is a break in the power spectrum at the angular scale $\Delta \theta \sim 1/\Gamma$.
\label{fig:mag} }
\end{figure}

\begin{figure*}
\epsscale{1.0}
\plotone{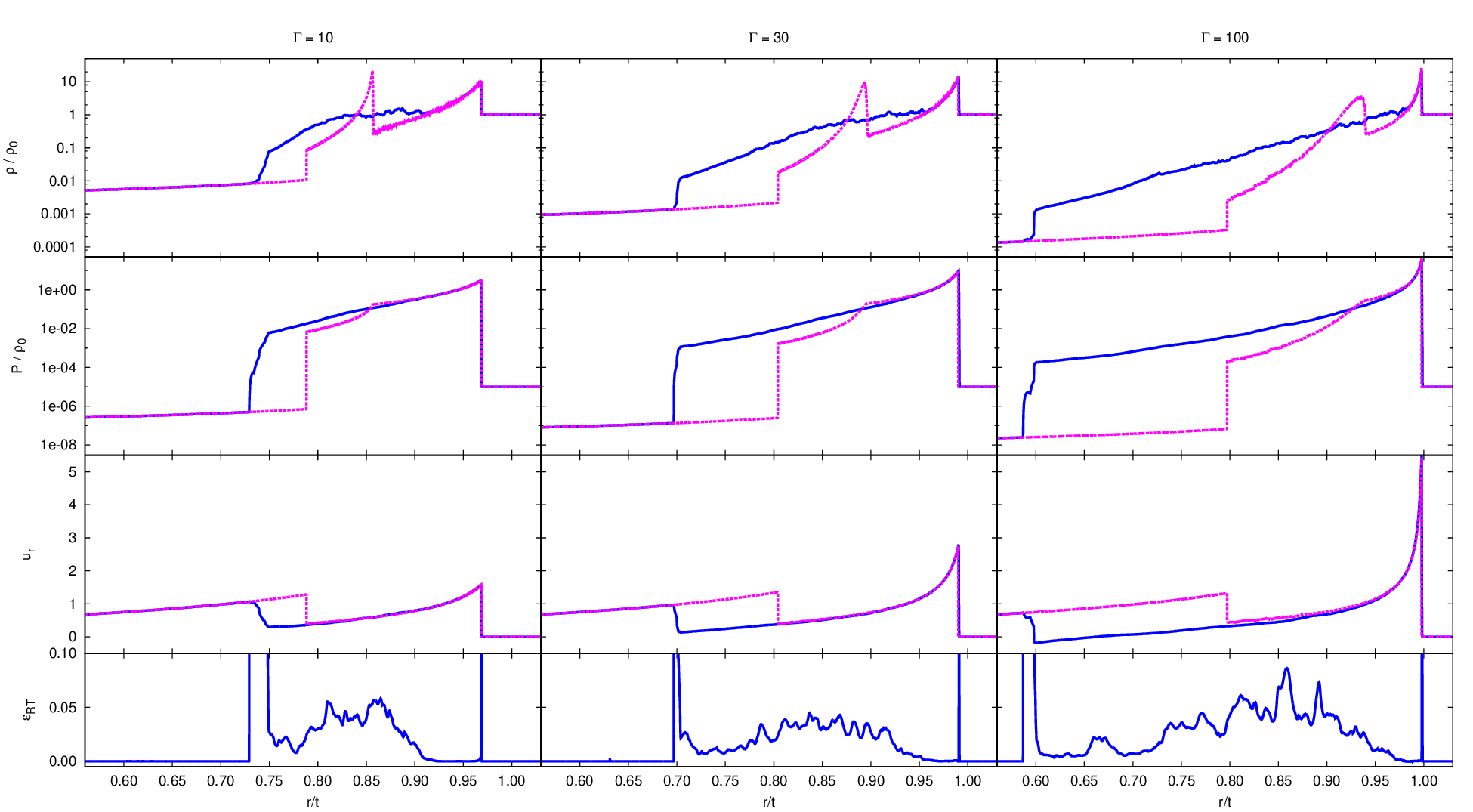}
\caption{The same calculation has been performed in 2D assuming axisymmetry, and in 1D assuming spherical symmetry.  The instability is present in two dimensions, and has a clear impact on the spherically averaged fields.  This is a snapshot at ${t = t_M}$ for $\Gamma = 10$, $30$, and $100$, comparing the 1D and 2D solutions.  The fields are spherically averaged.  The top three panels show a logscale profile in density, pressure and four velocity, showing clearly the position of the reverse shock, which is pushed back by the instability.  The contact discontinuity which is present in 1D is diffused away in 2D.  The bottom panel shows the turbulent energy fraction, $\epsilon_{RT}$ which is defined in \S \ref{sec:turb}.
\label{fig:1d2d} }
\end{figure*}

The calculation begins with a hot explosion with a given energy $E$ and mass $M$ with characteristic Lorentz factor ${\Gamma \equiv E/M}$, within a negligibly small radius, in a cold uniform-density medium with density $\rho_0$.  This specifies the problem completely, provided we have a definition for ``negligibly small radius".  The initial radius $R_0$ of the explosion must be small enough that the ``coasting phase" at $t \sim \Gamma R_0$ and the ``spreading phase" at $t \sim \Gamma^2 R_0$ occur long before the shell begins to decelerate, at the deceleration time
\begin{equation}
t_\gamma = \left( {M \over \Gamma \rho_0 }\right)^{1/3}.
\end{equation}
A detailed explanation of these different timescales is presented in \cite{ksp99}.  In this ``thin-shell" limit ($\Gamma^2 R_0 \ll t_\gamma$), the initial structure of the fireball has no effect on what happens after $t > t_\gamma$ (we have checked this empirically).  At $t \sim t_\gamma$, the shell decelerates, which causes the formation of the forward and reverse shocks.  This stage of the explosion is described by a self-similar solution which persists until the forward shock begins to sweep up more mass than is contained in ejecta, at $t = t_M$, where
\begin{equation}
t_M = ( M/\rho_0 )^{1/3}.
\end{equation}
During the phase of expansion between $t_\gamma$ and $t_M$, the RT instability is most significant.

To capture the first phases of evolution, we evolve the explosion in one dimension from the initial fireball until a time $t<t_\gamma$, when the mass in the forward shock is still a small fraction of the mass of the ejecta.  Then the code is switched to evolve the system in 2D, to capture the growth of the instability.  At this point, we also set the value of $X$ on either side of the blast wave ($X = 0$ behind and $X = 1$ in front).  The initial parameters are $E$, $M$, and $\rho_0$, but due to scale invariance, in this thin-shell limit the only parameter which must be varied is the energy per unit mass in the fireball, $E/M = \Gamma$.  Thus all the parameter space of this problem is reduced to this single variable.  In this study, we use values of $\Gamma = 10$, $30$, and $100$.  We have also performed calculations with $\Gamma = 300$ but the resolution was insufficient to correctly evolve the instability.  The $\Gamma > 100$ case will be presented in a future work.

In principle, the instability grows from some seed perturbations which break the spherical symmetry of the initial conditions.  However, we found that for weak enough perturbations the saturated instability does not depend (in a statistical sense) on how the perturbations are seeded.  This includes the limit where there are no explicit perturbations, but  the instability is seeded with numerical noise, which is extremely small but still large enough to break the symmetry.  In this work, we focus on the limiting case in which the seed vanishes, so we can focus completely on the effects of the instability by itself.  In reality, the instability can be seeded by significant density fluctuations in the circumburst medium, and the effects of these ``clumpy" fluctuations are very interesting, meriting a future study of their own.  In addition to our zero-seed calculations, we have also performed calculations which test the effect of seeding the instability with white noise density fluctuations.  This is discussed in appendix \ref{sec:seed}, though a proper study of the effects of a clumpy medium is outside the scope of this work.

\section{Results}
\label{sec:res}	

In Figure \ref{fig:times} we demonstrate how relativistic effects can modify the character of RT turbulence.  We show a snapshot at ${t = t_M}$ for various values of $\Gamma = E/M$, which is roughly a proxy for the initial Lorentz factor.  For larger Lorentz factors, the fastest growing modes can be found on smaller angular scales.  This has to do with causality; a light signal propagating along the shock front can only travel an angular distance $\Delta \theta = {2 \over 3 \gamma}$ for a blastwave decelerating with $\gamma \propto t^{-3/2}$ \citep{g00}.  Modes outside of this angular scale are ``frozen", i.e. unable to grow until the shock decelerates to a small enough Lorentz factor.  This is most explicitly shown in the power spectrum of the contact discontinuity, which is shown in Figure \ref{fig:mag}.  In this figure, a clear break is evident in the power spectrum at the angular scale $\Delta \theta \sim 1/\Gamma$.  The perturbation also propagates further both upstream and downstream for larger values of $\Gamma$.  For the turbulence moving back into the ejecta this is due to increased pressure and density gradients in the unstable region.  For the forward-propagating turbulence, it is due to the fact that the gas is compressed into a thin shell of width $\sim r/\Gamma^2$ behind the forward shock \citep{bm}, and the width of this shell decreases with increasing $\Gamma$.

\subsection{1D vs 2D}

The primary effects of the RT instability on the shock dynamics are displayed in Figure \ref{fig:1d2d}.  This is a snapshot of the shockwave at ${t = t_M}$ for ${\Gamma = 10}$, $30$ and $100$.  We show the radial profile and compare it with the same calculation performed in one dimension (1D) assuming spherical symmetry.  In the 1D case, we see the standard two-shock solution which has been extensively used to describe GRB phenomenology.

From this figure, the most obvious distinction between one and two dimensions is the complete disruption of the contact discontinuity in 2D, which of course is expected since it is the contact discontinuity which is unstable.  The second plainly visible distinction here is in the reverse shock.  In the 2D case, the instability has collided with the reverse shock, accelerating its propagation into the ejecta.  As a result, in the 2D case, the reverse shock propagates much further back than in 1D.  This apparently does not happen to the forward shock, as it remains undisturbed.  It is possible that this conclusion could be challenged by a three dimensional calculation, but we do not expect this to be the case, as the instability would dissipate more efficiently in 3D.

\begin{figure}
\epsscale{1.0}
\plotone{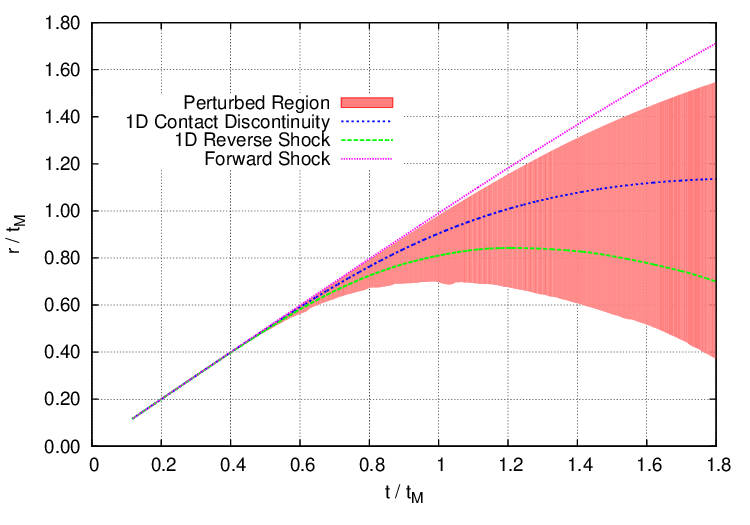}
\caption{Another comparison between the one-dimensional and two-dimensional versions of this experiment, for $\Gamma = 30$.  In this figure we plot the positions of the characteristic waves present within the shell.  The forward shock is unaffected by the instability, the contact discontinuity is sheared out into the large shaded region, and the reverse shock is pushed away from the forward shock.
\label{fig:shockpos} }
\end{figure}

Characteristic waves are shown in Figure \ref{fig:shockpos}, where we compare 1D and 2D shock dynamics.  Here again, we can see important effects of the instability:  What was a contact discontinuity becomes smeared over a larger region in 2D, overtaking the reverse shock but leaving the forward shock alone.  Characteristic positions of RT fingers are calculated using the passive scalar variable, $X$.  We find the grid zone with the largest value of $r$ for which $X<0.99$.  This value of $r$ we denote $r_{RT}$, the forward position of the RT fingers.

\begin{figure*}
\epsscale{1.0}
\plotone{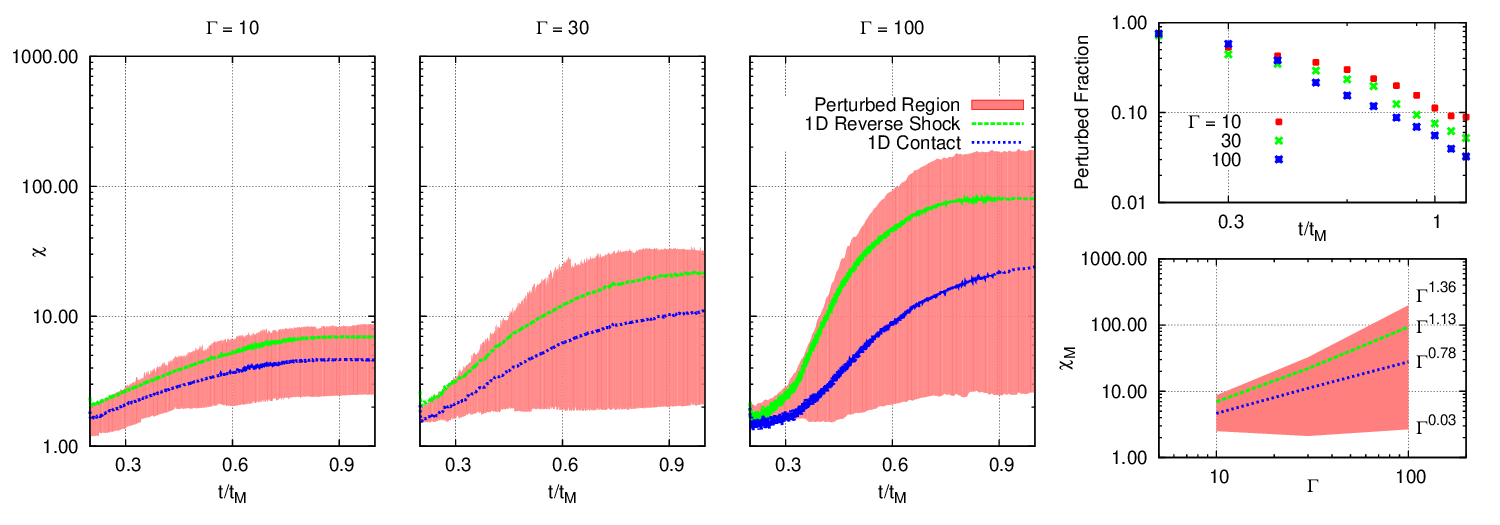}
\caption{  The fireball converts itself from a thin shell into the two-shock structure, and as $t \rightarrow t_M$ the characteristic waves obey self-similarity.  This remains true for the unstable 2D version of the problem.  Here we have plotted the evolution of the characteristic waves in terms of the self-similarity variable, $\chi$ (Eq. \ref{eqn:chi}).  Shortly after $t = t_M$, the characteristic waves fall behind the shock.  In the upper right panel we show the fraction of energy which is perturbed by the instability, of order 10\% at $t = t_M$.  In the lower right panel we show how the position of characteristic waves scales with Lorentz factor.
\label{fig:chi} }
\end{figure*}

Figure \ref{fig:shockpos} describes the characteristic waves in terms of the radial coordinate, r.  We can compare different values of $\Gamma$ on a similar footing if instead of r, we express our results in terms of the self-similarity parameter,
\begin{equation}
\chi = {1 - r/t \over 1 - r_{shock}/t}.
\label{eqn:chi}
\end{equation} 
We plot the value of $\chi$ for all characteristic waves in Figure \ref{fig:chi} for ${\Gamma = 10, 30,}$ and $100$.  As $t \rightarrow t_M$, the characteristic waves approach constant values of $\chi$.  From these three examples, it appears that the forward position of the RT fingers approaches a value of $\chi \sim 2.5$ which does not depend sensitively on $\Gamma$.  Since the forward shock is described by the self-similar Blandford-McKee solution, we can in principle use this value of $\chi$ to determine what fraction of the forward shock has been penetrated.  We calculate the ``penetration depth" of the RT fingers, defined as
\begin{equation}
d = {\int_{r < r_{RT}} \tau d^3x \over \int_{\text{all space}} \tau d^3x}
\end{equation}
where $\tau = T^{0 0} - \rho u^0$ is the (non-rest-mass) energy density and $r_{RT}$ denotes the forward-most radial position of the RT fingers.  A straightforward calculation using the Blandford-McKee solution yields $d = \chi^{-17/12} \approx 0.27$ of the forward shock is affected by the instability, though this assumes an ultrarelativistic forward shock, and that the forward shock carries most of the explosion energy, neglecting the energy in the ejecta.  We can avoid these assumptions by calculating this directly from the data.  In this case, we determine how far the instability has penetrated into the shock by calculating the ratio
\begin{equation}
d = {\int (1-\left<X\right>) \left<\tau\right> r^2 dr \over \int \left<\tau\right> r^2 dr}
\end{equation}
where the brackets denote an average over the polar angle, $\theta$.  We find in a direct calculation that the penetration depth is a function of time, which is close to 1/3 when $t \sim 0.3 t_M$ but decreases with time, and is of order $10\%$ around $t = t_M$ (upper-right panel of Figure \ref{fig:chi}).  At $t = t_M$, $d = 11\%$ for $\Gamma=10$, $8\%$ for $\Gamma=30$, and $6\%$ for $\Gamma=100$.  This is the percentage which is turbulent and therefore magnetized by RT (we calculate the magnetization in \S \ref{sec:turb}).  After $t = t_M$, the shock moves away from the fingers quickly and they have no further impact on the forward shock dynamics.  This will be discussed in more detail in the next section.

\subsection{Turbulence}
\label{sec:turb}

It will be useful to measure the kinetic energy of large-scale turbulent fluctuations, since that will provide insight into the turbulent cascade which would take place in a highly-resolved 3D calculation.  We characterize this kinetic energy as a fraction of thermal energy:
\begin{equation}
\epsilon_{RT} = { \left<\delta KE\right> \over \left<TE\right> }
\end{equation}
where $\delta KE$ is the kinetic energy of fluctuations and $TE = 3P$ is the thermal energy density.  We have found a simple way of calculating this quantity, the details of which can be found in appendix \ref{sec:eps}.  In our analysis, we have converted 2D data into 1D data by averaging in a conservative way.  That is, for a given spherical shell, the total mass, energy, and momentum of that shell are calculated, and assuming the fields are uniform throughout the shell, the density, pressure, and four velocity are calculated from the energy, momentum, and mass of the shell (this is how the middle panels of Figure \ref{fig:1d2d} were generated).  We note that this averaging process typically results in a slightly larger internal energy than would result from arithmetically averaging the pressure by volume on a spherical shell.  This is because when the momentum of a shell is calculated, there is cancellation due to the variations in velocity with angle, and this ``lost" kinetic energy is interpreted as internal energy in the averaged quantities.

In other words, the excess of internal energy from averaging conservatively over angle is exactly the turbulent kinetic energy contained in that spherical shell.  This provides a very simple means of calculating the fraction of energy in turbulent fluctuations:
\begin{equation}
\epsilon_{RT} = {\Delta w \over \left<3 P\right>}
\end{equation}
where $w = \rho + 3P$ is the internal energy density, and $\Delta w$ is the difference between ``conservatively averaged" internal energy density and volume-averaged internal energy density.  We perform a more careful derivation of this formula in appendix \ref{sec:eps}.

\begin{figure}
\epsscale{1.0}
\plotone{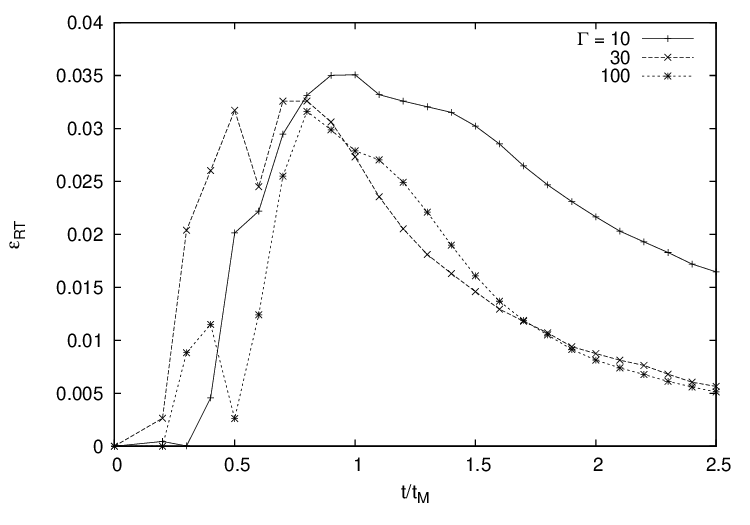}
\caption{ Volume-Averaged turbulent kinetic energy fraction $\bar \epsilon_{RT}$ in the affected region, as a function of time.  After time $t > t_M$, the turbulence is no longer driven by RT, and $\epsilon_{RT}$ decreases with time.
\label{fig:epstimes} }
\end{figure}

We have plotted a snapshot of $\epsilon_{RT}$ in the lower panels of Figure \ref{fig:1d2d}.  There is clearly variation with radius, but in this snapshot it roughly peaks around ${\epsilon_{RT} \sim 0.05}$ and has an average value of ${\bar \epsilon_{RT} \sim 0.03}$ for each value of $\Gamma$.  For the three cases studied, this turbulent fraction peaks at time $t \sim t_M$ with a peak average of roughly 0.03, also roughly independent of $\Gamma$.  We show this in Figure \ref{fig:epstimes}.

Although we have not included magnetic fields in our numerical calculations, it is possible to infer that magnetic energy density will be in equipartition with the local turbulent kinetic energy density behind the forward shock.  Equipartition is established by turbulent dynamo processes, whose activity is guaranteed by the presence of pre-existing magnetic fluctuations (however small), extremely large Reynolds and magnetic Prandtl number of the fluid, and continuous injection of turbulent kinetic energy by the RT instability.  First, by repetitive stetching and folding of frozen-in magnetic field lines, kinematic small-scale turbulent dynamo drives exponential growth of the pre-existing magnetic field at a rate comparable to the turnover of smallest eddies \citep{k68,m78}.  This rate is extremely fast compared to outer-scale eddy turn-over times due to the large Reynolds numbers involved.  The kinematic process terminates when the energy in viscous scale magnetic fluctuations balances kinetic energy of viscous scale eddies. Magnetic field amplification then continues via nonlinear small-scale dynamo process \citep{s04}.  This process is characterized by linear growth of the overall magnetic energy, and the formation of magnetic field fluctuations at progressively larger scales.  It terminates when magnetic fluctuations exist in scale-by-scale equipartition up to the scale of the RT fingers. Numerical simulations in the non-relativistic \citep{hbd03,b12} and relativistic \citep{zrakey2} cases indicate that in the limit of large Reynolds number, non-linear small-scale turbulent dynamo saturates universally after several large-scale turnover times, notwithstanding effects due to high Mach number \citep{f10}.  Therefore, if magnetic fields and the turbulent sub-scales could be included in our calculations, we anticipate magnetic fields in kinetic equipartition with the RT fingers over the lifetime of the instability.  ``Equipartition" here means that the ratio of kinetic to magnetic energy is of order unity.  The aforementioned numerical studies have reported end-state turbulent dynamo saturation with magnetic energy at between 30\% and 60\% of the turbulent kinetic energy density:
\begin{equation}
\epsilon_B / \epsilon_{RT} \approx 0.3 - 0.6
\end{equation}
The implied $\epsilon_B$ for our study is thus
\begin{equation}
\epsilon_B \sim 10^{-2},
\end{equation}
within a factor of two.

This value is strong enough to produce observed afterglow synchrotron emission \citep{pk02}, but it should be noted that magnetic fields are only present in the fraction of the blastwave which is penetrated by RT turbulence (of order 10\%, as discussed in the previous section).

\subsection{Afterglow Emission}

We have calculated light curves in one and two dimensions, in order to explore the observational consequences of the instability.  Relativistic Doppler effects are responsible for the hard gamma-ray and X-ray afterglow emission, but the reverse shock generally has a much lower Lorentz factor than the forward shock, so it cannot compete with the forward shock at high frequencies.  At lower frequencies (optical, microwave, radio) the flux from the forward and reverse shocks can be comparable, so in these wavelengths we search for observational signatures of the RT instability.

By implementing a simple model for synchrotron emission directly into the hydrodynamical evolution, we calculate a microwave light curve for our $\Gamma = 30$ model.  We calculate the flux from a given computational zone at a given observer time from the formula:
\begin{equation}
F \propto { \rho B \over \gamma^2 (1-v_z)^2 } Q(\nu)
\end{equation}
where the magnetic energy is assumed to be proportional to the thermal energy,
\begin{equation}
B \propto \sqrt{P}
\end{equation}
and frequency dependence assumes ``slow cooling"
\begin{equation}
Q(\nu) = \left\{ \begin{array}{ll}
(\nu/\nu_m)^{1/3}     & \nu < \nu_m \\
(\nu/\nu_m)^{(1-p)/2} & \nu > \nu_m
\end{array} \right.
\end{equation}
where we choose p = 2.5, and where $\nu_m$ is the characteristic synchrotron frequency,
\begin{equation}
\nu_m \propto  B (P / \rho)^2
\end{equation}
This is a simplified version of the emission model in \cite{hendrik}.

\begin{figure}
\epsscale{1.0}
\plotone{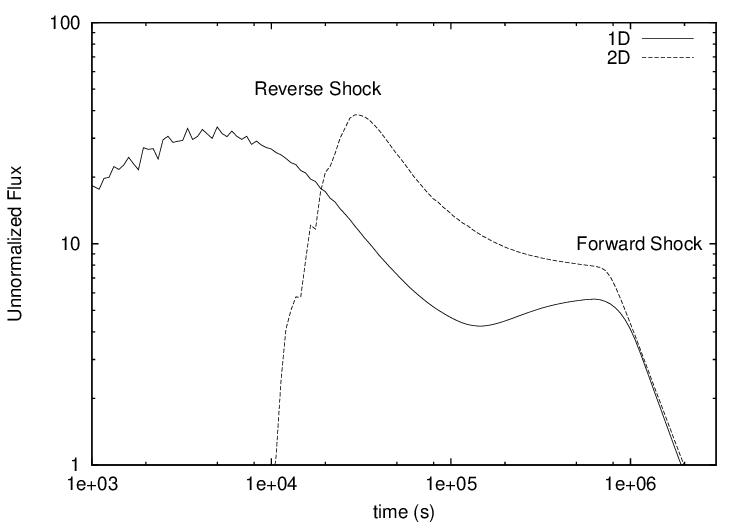}
\caption{ The reverse shock is directly observed in the early afterglow emission.  We plot microwave light curves for $\Gamma = 30$ for the 1D (stable) and 2D (unstable) versions of the problem.  The second peak is due to the forward shock, which is unaffected by the instability.  The first peak is emission from the reverse shock.  This flash occurs at a later time and the peak is narrower.
\label{fig:lightcurve} }
\end{figure}

We show the light curve in Figure \ref{fig:lightcurve} for $\Gamma = 30$.  To single out signatures of RT, we compare 1D and 2D results.  The forward shock emission is unaffected, as the forward shock is not perturbed by the instability.  The primary impact on the reverse shock is to modify its Lorentz factor, and its value of $\nu_m$ at the peak reverse shock emission time, which changes the time this peak enters the observed frequency band. The peak magnitude is not diminished significantly, but the peak occurs at a later time in 2D than in 1D.  This light curve is intended to qualitatively demonstrate the impact of RT; a more careful calculation of light curves with a radiation transfer code will be carried out in a future work.

\section{Summary}
\label{sec:sum}

We have computed the development and saturation of Rayleigh-Taylor instability in relativistic fireballs using a moving computational mesh.  The instability does not perturb the forward shock, but it completely disrupts the contact discontinuity between the ejecta and the surrounding medium, and it significantly modifies the dynamics of the reverse shock, which causes reverse shock emission to peak at a later time.  The relativistic shear flow generated by RT fingers may also provide a site for cosmic ray acceleration.

The Rayleigh-Taylor instability is effective in stirring up turbulence; we find that turbulent kinetic energy reaches about 3\% of thermal energy wherever the turbulence is present, which should rapidly amplify magnetic fields up to 1-2\% of thermal by the operation of small-scale turbulent dynamo (The latter percentage was not directly taken from our numerical calculation, but it was inferred from the results of various MHD turbulence studies).  The turbulence is present in a region behind the forward shock, but RT fingers attempt to reach out towards this shock, penetrating of order $10\%$ of the way towards the shock front ($11\%$ for $\Gamma=10$, $8\%$ for $\Gamma=30$, and $6\%$ for $\Gamma=100$).  This motivates 3D calculations, as these numbers might change when 3D turbulence effects are taken into account.  Some of these results might also be modified by a coherent seed field in the form of a clumpy medium.  In fact, it is possible that such a seed field might be enough to drive RT fingers to make contact with the forward shock.

This emphasizes the importance of accounting for fluid instabilities when considering the dynamics and observational signatures of gamma ray bursts.  Figure \ref{fig:1d2d} gives some hope that an effective 1D description of the dynamics may be possible, akin to the convection model of \cite{g73}; adding some kind of diffusive term in 1D to shocked fluid elements in the ejecta might reproduce physical profiles close to those in Figure \ref{fig:1d2d}.

This motivates a more comprehensive study which includes more general circumburst profiles, larger Lorentz factors (converged results with $\Gamma \sim 300$ or higher are much more computationally intensive but certainly possible), and potentially 3D turbulence effects.  Because of scale invariance, the only free parameter is $\Gamma$, the characteristic Lorentz factor.  The dynamics and synchrotron emission both can be re-scaled to arbitrary explosion energy and circumburst density, which means that a suite of calculations for different values of $\Gamma$ will completely cover the parameter space of all such explosions \citep{scaling, granot}.  This motivates us to implement a detailed synchrotron emission model \citep{hendrik}, to perform a comprehensive study of emission from GRB outflows.  All of this is planned as future work.

\acknowledgments
This research was supported in part by NASA through grants NNX10AF62G and NNX11AE05G issued through the Astrophysics Theory Program and by the NSF through grant AST-1009863.  

Resources supporting this work were provided by the NASA High-End Computing (HEC) Program through the NASA Advanced Supercomputing (NAS) Division at Ames Research Center.  We are grateful to Andrei Gruzinov, and Hendrik van Eerten for helpful comments and discussions.  We thank Jonathan Zrake for illuminating discussions about magnetic field amplification via small scale turbulent dynamo.  We also would like to thank the anonymous referee for his or her extremely thorough review.

\begin{appendix}
\section{Resolution Study}

\begin{figure}
\epsscale{1.0}
\plotone{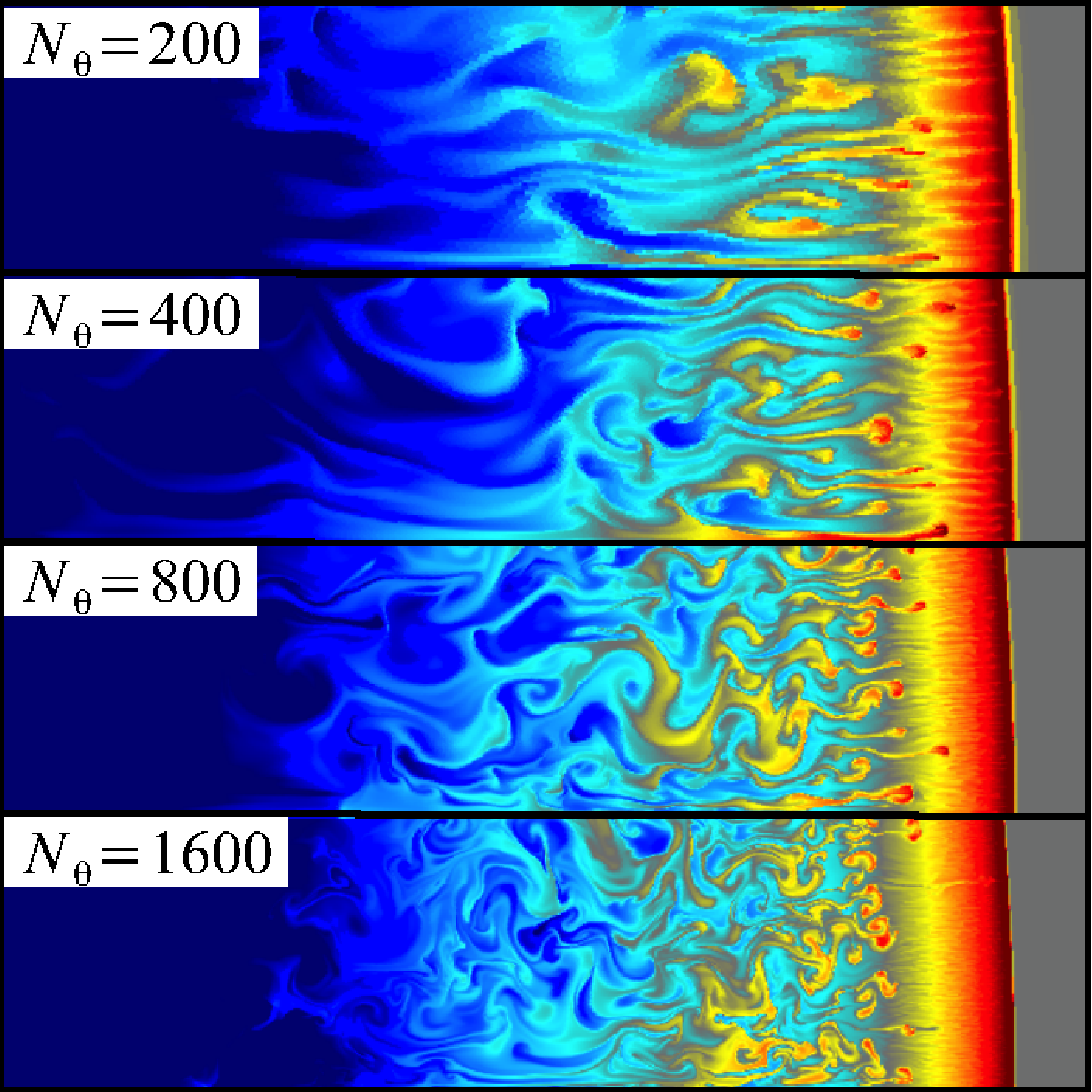}
\caption{ A qualitative picture of resolution dependence for $\Gamma = 30$ at $t = t_M$ (Color bar is logarithm of density, the same scaling as in Figure \ref{fig:times}).  Even at low resolution, the shock dynamics are reasonably well-resolved.  Resolving the turbulence, however, requires higher resolution.  Somewhere between $400$ and $800$ cells appears necessary to resolve the outer scale of turbulence properly.
\label{fig:resstudy} }
\end{figure}

\begin{figure}
\epsscale{1.0}
\plotone{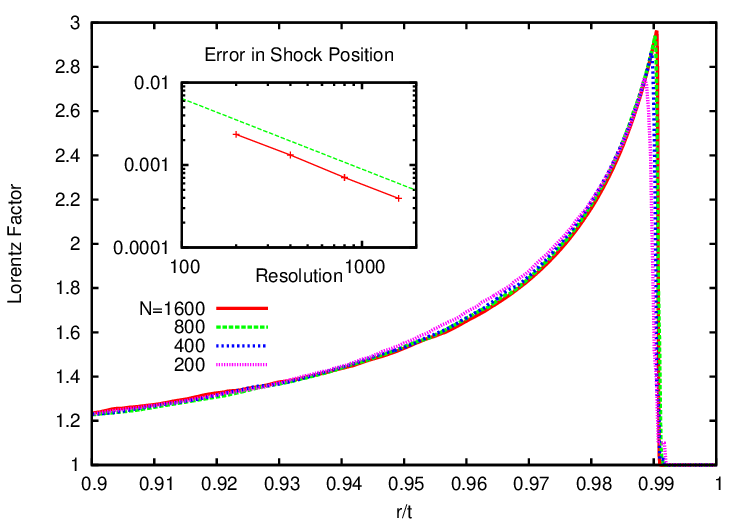}
\caption{ Convergence test of the forward shock position at $t = t_M$.  Shock dynamics are extremely well captured; the convergence rate is less than first order (the slope of the line in the inset is 0.85) but the error bars on the forward shock position are very small: $\Delta r / r \sim 4 \times 10^{-4}$.
\label{fig:resshock} }
\end{figure}

\begin{figure}
\epsscale{1.0}
\plotone{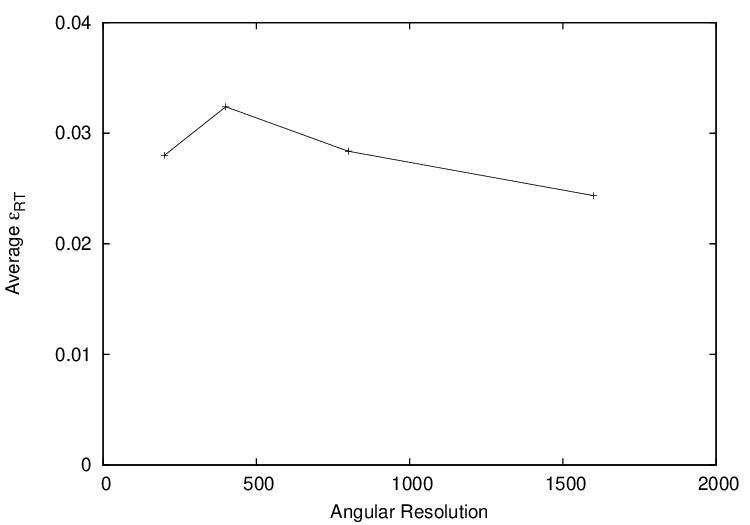}
\caption{ Convergence test of the averaged turbulent energy fraction, $\bar \epsilon_{RT}$, at $t = t_M$.  Convergence is slow for the turbulence; the value of $\epsilon_{RT}$ varies by $\sim 30\%$ as we vary the resolution.  This establishes that our value of $\epsilon_{RT}$ is reasonably accurate, but probably not to within 10\%.
\label{fig:resturb} }
\end{figure}

We have performed a series of calculations for the $\Gamma = 30$ example at various resolutions, to gather a sense of how large the error bars are in our results.  The calculations which derived our main results used $N_{\theta} = 1600$ zones, an angular resolution of $\Delta \theta = 1.2 \times 10^{-4}$, and an equivalent (variable) radial spacing $\Delta r / r \sim \Delta \theta$.  We find that this resolution is plenty for computing the basic shock dynamics, but that turbulence is more difficult to resolve accurately.

In Figure \ref{fig:resstudy}, we show the qualitative dependence of the solution on the resolution, for $N_{\theta} = 200$, $400$, $800$, and $1600$, with the radial resolution scaling as $\Delta r / r \sim \Delta \theta$.  The shock dynamics are roughly the same, but qualitatively it seems that the turbulence is only well-resolved in the $N_{\theta} = 800$ and $1600$ cases.  To determine the accuracy of shock dynamics, we look at the profile of the forward shock at $t = t_M$ in Figure \ref{fig:resshock}.  Even the low-resolution calculations fit the shock reasonably well, and it is clear from the picture that the profile of Lorentz factor is converging.  To be more quantitative, we calculate the error in the position of the forward shock at $t = t_M$, and find the convergence to be slightly lower than first order (we calculate an exponent of 0.85 for the order of convergence in this test).  The error in the forward shock position $\Delta r / r \sim 4 \times 10^{-4}$ at our highest resolution.

Turbulence is more difficult to capture in this problem, because it requires resolving enough scales that the outer scale of turbulence is largely ignorant of the inner scale, given by the grid spacing.  The turbulent kinetic energy fraction $\epsilon_{RT}$ is much more sensitive than the shock position to variation of the resolution.  We examine how this number (averaged over the perturbed region at $t=t_M$) varies with resolution in Figure \ref{fig:resturb}.  We have found that $\epsilon_{RT}$ is roughly $3 \times 10^{-2}$, but this number fluctuates by $\sim 30\%$ as we vary the resolution.  Hence, the relative error bars on $\epsilon_{RT}$ are much larger than that of the shock positions.  However, these errors (at the level of $30\%$) are still smaller than uncertainties in magnetic field amplification, and probably smaller than uncertainties in unmodeled physics, like three-dimensional effects.

\section{Calculation of Turbulent Kinetic Energy}
\label{sec:eps}
In this section we formally demonstrate that our diagnostic $\epsilon_{RT}$ measures the kinetic energy of turbulent fluctuations as a fraction of the thermal energy.  We first specifically define all quantities, demonstrate that $\epsilon_{RT}$ is a Lorentz-invariant quantity, then we calculate it in a convenient reference frame.
\subsection{Definitions}
First, we define some region being averaged over, $\Omega$.  This could be any spacetime region, though in our case it is a thin spherical shell.  Call the volume element of this region $d\Omega$.  In our analysis, we use two kinds of averages over $\Omega$.  First, there is the simple volume average:
\begin{equation}
\left<X\right>_{Vol} = { \int X d\Omega \over \int d\Omega }
\end{equation}
Then, there is what we call the ``conservative average".  To determine conservative averages, we first calculate the volume average of the conserved quantities, $\rho u^0$, and $T^{0 \mu}$.  Then, making the assumption that the conserved quantities are uniformly distributed over $\Omega$, we calculate the primitive variables $\rho$, $P$, and $u^{\mu}$.  We define these calculated primitive variables to be the ``Conservatively averaged proper density" and so on.  Any quantity we calculate using these primitive variables we define to be ``conservatively averaged".  In particular, for conserved quantities, the conservative average just equals the volume average:
\begin{equation}
\left< \tau \right>_{Cons} = \left< \tau \right>_{Vol}.
\end{equation}
We define the quantity $\epsilon_{RT}$ in the following manner.  Define $w \equiv \rho + 3P$, the proper internal energy.  This quantity can be conservatively averaged or volume averaged.  Define the difference $\Delta w \equiv \left<\rho\right>_{Cons} + 3\left<P\right>_{Cons} - \left<\rho\right>_{Vol} - 3\left<P\right>_{Vol}$.  Then define the turbulent energy fraction:
\begin{equation}
\epsilon_{RT} \equiv { {\Delta w} \over 3\left<P\right>_{Vol} }.
\end{equation}
\subsection{Lorentz Invariance}

The fact that $\epsilon_{RT}$ is a Lorentz invariant is not essential to its utility; we are only demonstrating it because we use this fact in the next subsection.  We should also state specifically what we mean by ``Lorentz invariance".  It is clear that upon performing a Lorentz transformation, the region $\Omega$ can change substantially; in particular, if it is a spatial volume in one frame, it becomes a spacetime hypervolume in another reference frame.  This fact does not concern us; for our purposes ``Lorentz Invariance" simply means that if the integration region is transformed properly from one frame to another, the calculation of $\epsilon_{RT}$ in one frame over $\Omega$ yields the same number as if it was calculated in a different frame over $\Omega'$, the suitably Lorentz-transformed spacetime region.

The argument that $\epsilon_{RT}$ is independent of reference frame simply follows from the fact that $\rho$ and $P$ are proper quantities.  Because these primitive variables are independent of reference frame, and because both averaging processes are also independent of reference frame (all volume elements $d\Omega$ are transformed by the same factor under Lorentz boosts), $\epsilon_{RT}$ is also a Lorentz-invariant.

\subsection{Calculation in Center of Momentum Frame}

\begin{figure}
\epsscale{1.0}
\plotone{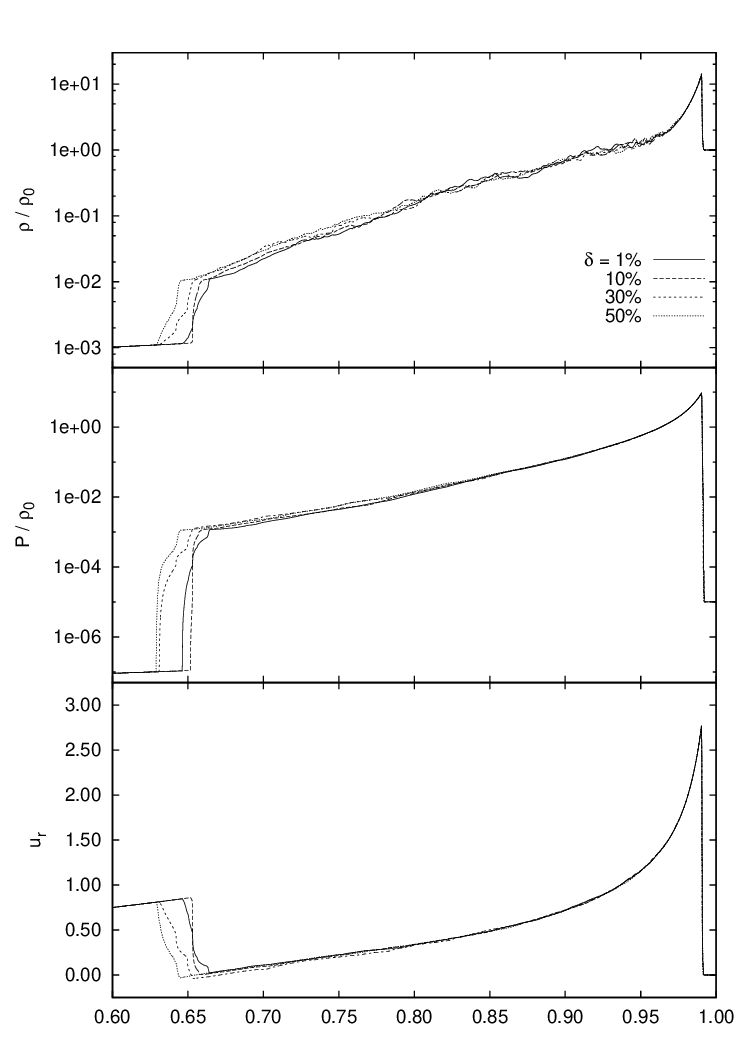}
\caption{ The instability is seeded with white-noise perturbations in the density field.  Even with root-mean-square fluctuations as large as $50\%$ of $\rho_0$, the spherically averaged fields are only weakly affected by the perturbations.  For white noise, the final solution is essentially independent of the seed field (in a mean-field sense).
\label{fig:seed_1d} }
\end{figure}

All of the following calculations are performed in the center-of-momentum frame of $\Omega$.  In other words, we work in a frame in which
\begin{equation}
\left<\vec S\right>_{Vol} = \left<\vec S\right>_{Cons} = 0,
\end{equation}
where $S^i = T^{0 i}$ is the momentum density.  If the momentum is zero, this guarantees that the conservatively-averaged velocity is zero:
\begin{equation}
\left<\vec v\right>_{Cons} = 0.
\end{equation}
Therefore, the conservatively averaged energy in this frame is just given by the $(0,0)$ component of Equation \ref{eqn:stress} when we set the velocity to zero:
\begin{equation}
\left< T^{0 0} \right>_{Cons} = \left<\rho\right>_{Cons} + \left<3 P\right>_{Cons}.
\end{equation}
Note that the conservatively averaged energy density should be equal to the volume-averaged energy density, since energy is a conserved quantity.  The volume-averaged energy density is:
\begin{equation}
\left< T^{0 0} \right>_{Vol} = \left<\rho\right>_{Vol} + \left<3 P\right>_{Vol} + \left<\delta KE\right>_{Vol}.
\end{equation}
We could interpret this equation as a definition of $\delta KE$; that is, it is exactly the energy which is not thermal energy or rest mass energy.  The bulk kinetic energy in this frame is zero, so all kinetic energy comes from fluctuations.  We should point out that this interpretation assumes that the region $\Omega$ is small and that the large-scale mean fields do not vary significantly over $\Omega$, so that all variations can be considered fluctuations about the mean.  The spherical shell we use for $\Omega$ is extremely thin, roughly the grid spacing, which means this interpretation is generally valid.  If $\Omega$ crosses a shock front, the interpretation breaks down.  It is worth pointing out that this definition of $\delta KE$ is consistent with definitions used in local numerical studies of driven turbulence, because these studies generally measure kinetic energy in the center of momentum frame.

Now, by equating conservatively averaged energy with volume-averaged energy,
\begin{equation}
\left< T^{0 0} \right>_{Cons} = \left< T^{0 0} \right>_{Vol}
\end{equation}
we immediately find that the kinetic energy in fluctuations is given by the internal energy difference defined before:
\begin{equation}
\left<\delta KE\right>_{Vol} = \Delta w.
\end{equation}
So, we have our intended result:
\begin{equation}
\epsilon_{RT} = { \left<\delta KE\right>_{Vol} \over \left<TE\right>_{Vol} } = { \Delta w \over 3\left<P\right>_{Vol} }
\end{equation}
where $TE = 3P$ is the thermal energy.  Additionally, we note that this formula simplifies in the nonrelativistic case and the ultrarelativistic case.  In both cases,
\begin{equation}
\epsilon_{RT} \rightarrow { \Delta P \over \left< P \right>_{Vol} }
\end{equation}

where $\Delta P = \left<P\right>_{Cons} - \left<P\right>_{Vol}$.  In the ultrarelativistic limit, this is because $P \gg \rho$.  In the nonrelativistic case, it is because density is a conserved variable and so ${ \left<\rho\right>_{Cons} = \left<\rho\right>_{Vol} }$.

\section{Effect of Seeding the Instability}
\label{sec:seed}

We stated in \S \ref{sec:ics} that our results are to be considered in the limiting case in which seed perturbations vanish.  In this section we study the effect of white noise perturbations in the form of a fluctuating density profile,
\begin{equation}
\rho = \rho_0 + \delta \rho(\vec x),
\end{equation}
which we parameterize in terms of root-mean-square fluctuations:
\begin{equation}
\delta = {1 \over \rho_0} \sqrt{ \int \delta \rho^2 dV }.
\end{equation}
We performed calculations with $\delta = 1\%$, $10\%$, $30\%$, and $50\%$, and found very little statistical dependence on $\delta$.  Naturally, the positions of individual eddies are extremely sensitive to perturbations, but the mean-field quantities, like the spherically averaged fields, are nearly independent of $\delta$ (Figure \ref{fig:seed_1d}).

This demonstrates that the limit $\delta \rightarrow 0$ is meaningful and using numerical noise to seed the instability gives the same evolution (in a statistical sense) as using a small white-noise seed field.

A coherent seed field, on the other hand, may be enough to qualitatively influence the solution.  Circumburst profiles are likely to be clumpy, and this may play an important role in the dynamics \citep{rr03,mnz07}.  The influence of such perturbations is an important topic which is outside the scope of the current work.

\end{appendix}

{}

\end{document}